\documentclass[twocolumn,showpacs,preprintnumbers,amsmath,amssymb]{revtex4}
\usepackage{graphicx}% Include figure files
\usepackage{bm}% bold math

\begin{document}
\title{Monopole and Berry Phase in Momentum Space \\
in Noncommutative Quantum
Mechanics}
\author{Alain B\'{e}rard and Herv\'{e} Mohrbach}
\affiliation{Laboratoire de Physique Mol\'eculaire et des
Collisions, Institut de Physique, Technop\^ole 2000, 57078 Metz,
France}

\date{\today}% It is always \today, today,
\begin{abstract}
To build genuine generators of the rotations group in
noncommutative quantum mechanics, we show that it is necessary to
extend the noncommutative parameter $\theta $ to a field operator,
which one proves to be only momentum dependent. We find
consequently that this field must be obligatorily a dual Dirac
monopole in momentum space. Recent experiments in the context of 
the anomalous Hall effect provide evidence for a monopole in the 
crystal momentum space. We suggest a connection between the 
noncommutative field and the Berry curvature in momentum space 
which is at the origine of the anomalous Hall effect. 
\end{abstract}

\pacs{14.80.Hv}% PACS, the Physics and Astronomy
\maketitle

A natural generalization of Quantum Mechanics involving
noncommutative space time coordinates was originally introduced by
Snyder \cite{SNYDER} as a short distance regularization to improve
the problem of infinite self energies inherent in a Quantum Field
Theory. Due to the advent of the renormalization theory this idea
was not very popular until A. Connes \cite {CONNES} analyzed Yang
Mills theories on noncommutative space.\ More recently a
correspondence between a non-commutative gauge theory and a
conventional gauge theory was introduced by Seiberg and Witten
\cite{SEIBERG}. Noncommutative gauge theories were also found as
being naturally related to string and M-theory \cite{KONECHNY}.

In this framework an antisymmetric $\theta ^{\mu \nu }$ parameter
usually taken to be constant \cite{DERIGLASOV,ROMERO} is
introduced in the commutation relation of the coordinates in the
space time manifold $\ \left[ x^{\mu },x^{\upsilon }\right]
=i\theta ^{\mu \nu }.$ This relation leads to the violation of the
Lorentz symmetry, a possibility which is intensively studied
theoretically and experimentally \cite{KOSTELECKY}. Applications
of noncommutative theories were also found in condensed matter
physics, for instance in the Quantum Hall effect \cite{BELLISSARD,BELISSARD1}
and the non-commutative Landau problem \cite{GAMBOA,JACKIW1,HORVATHY}
i.e., a quantum particle in the non-commutative plane, coupled to
a constant magnetic field with a constant selected $\theta $
parameter as usual.

In this letter, we generalize the quantum mechanics in non-commutative
geometry by promoting the $\theta $ parameter with a new field obeying its
own field equations. Note that some authors, (for example \cite{DAS}) introduced 
a position dependent $\theta $ field using a Kontsevich product \cite{KONTSEVICH} 
in the study of gauge theory. Contrary to these approaches we find that the $\theta $
field must be momentum dependent.

The physical motivations of our work are twofold:

(i) For a constant $\theta $ field, we show that a quantum particle in a
harmonic potential has a behavior similar to a particle in a constant
magnetic field $\theta $ in standard quantum mechanics, since a paramagnetic
term appears in the Hamiltonian. Moreover the particle in the presence of
the $\theta $ field acquires an effective dual mass in the same way that an
electron moving in a periodic potential in solid state physics. Thus it is
legitimate to interpret this field like a field having properties of the
vacuum. In this context it is natural to extend the theory to a non-constant
field. This proposal is strongly enforced by the lack of rotation generators
in noncommutative space with a constant $\theta $ parameter, i.e. the
angular momentum does not satisfy the usual angular momentum algebra. We
then show that this $\theta $ field is only momentum dependent and that the
requirement of the angular momentum algebra, that is the existence of an
angular momentum, necessarily imposes a dual Dirac monopole in momentum
space field configuration. Thereafter we will intensely use the concept of
duality between the quantities defined in momentum space compared with those
defined in the position space. 

(ii) The second motivation comes for recent theoretical works
\cite{ONODA} concerning the anomalous Hall effect in
two-dimensional ferromagnets predicting topological singularity in
the Brillouin zone, but especially very recent experiments carried
out in the same context \cite{FANG} where a monopole in the
crystal momentum space seems to have been discovered. This monopole being a
singular configuration of the Berry curvature it appears naturally in the 
expression of the Hall conductivity \cite{THOULESS}. We
will consider this framework as a physical realization
of our more general theory, where the Berry curvature corresponds
to our $\theta (p)$ field.

Consider a quantum particle of mass $m$ whose coordinates satisfy the
deformed Heisenberg algebra
\[
\left[ x^{i},x^{j}\right] =i\hbar q_{\theta }\theta ^{ij}(\mathbf{x},\mathbf{%
p})\text{ ,}
\]
\[
\left[ x^{i},p^{j}\right] =i\hbar \delta ^{ij}\text{ ,}
\]
\[
\text{ }\left[ p^{i},p^{j}\right] =0,
\]
where $\theta $ is a field which is a priori position and momentum dependent
and $q_{\theta }$ is a charge characterizing the intensity of the 
interaction of the particle and the $\theta $ field. Note that we do not 
consider any external magnetic field in this work, but its taking into account 
does not pose a problem.
It is well known that these commutation relations can be obtained from the deformation of the
Poisson algebra of classical observable with a provided Weyl-Wigner-Moyal
product \cite{MOYAL} expanded at the first order in $\theta $.

The following Jacobi identity
\begin{equation}
\left[ p^{i},\left[ x^{j},x^{k}\right] \right] +\left[ x^{j},\left[
x^{k},p^{i}\right] \right] +\left[ x^{k},\left[ p^{i},x^{j}\right] \right]
=0,
\end{equation}
implies the important property that the $\theta $ field is position
independent
\begin{equation}
\theta ^{jk}=\theta ^{jk}(\mathbf{p}).
\end{equation}
Then one can see the $\theta $ field like a dual of a magnetic field and $%
q_{\theta }$ like a dual of an electric charge. The fact that the field is
homogeneous in space is an essential property for the vacuum. In addition,
one easily see that a particle in this field moves freely, that is, the
vacuum field does not act on the motion of the particle in the absence of an
external potential. The effect of the $\theta $ field is manifest only in
presence of a position dependent potential.

To look further at the properties of the $\theta $ field consider the other
Jacobi identity

\begin{equation}
\left[ x^{i},\left[ x^{j},x^{k}\right] \right] +\left[ x^{j},\left[
x^{k},x^{i}\right] \right] +\left[ x^{k},\left[ x^{i},x^{j}\right] \right]
=0,  \label{jacob}
\end{equation}
giving the equation of motion of the field
\begin{equation}
\frac{\partial \theta ^{jk}(\mathbf{p})}{\partial p^{i}}+\frac{\partial
\theta ^{ki}(\mathbf{p})}{\partial p^{j}}+\frac{\partial \theta ^{ij}(%
\mathbf{p})}{\partial p^{k}}=0,  \label{div}
\end{equation}
which is the dual equation of the Maxwell equation $\ div\overrightarrow{B}%
=0.$ As we will see later, equation (\ref{div}) is not satisfied in the
presence of a monopole and this will have important consequences.

Now consider the position transformation
\begin{equation}
X^{i}=x^{i}+q_{\theta }a_{\theta }^{i}(\mathbf{x},\mathbf{p}),
\label{position}
\end{equation}
where $a_{\theta }$ is a priori position and momentum dependent, that
restores the usual canonical Heisenberg algebra
\[
\left[ X^{i},X^{j}\right] =0,
\]
\[
\text{ }\left[ X^{i},p^{j}\right] =i\hbar \delta ^{ij}{\text ,}
\]
\[
\left[ p^{i},p^{j}\right] =0{\text .}
\]
The second commutation relation implies that $a_{\theta }$ is position
independent, while the commutation relation of the positions leads to the
following expression of $\theta $ in terms of the dual gauge field $%
a_{\theta }$%
\begin{equation}
\theta ^{ij}(\mathbf{p})=\frac{\partial a_{\theta }^{i}(\mathbf{p})}{%
\partial p^{j}}-\frac{\partial a_{\theta }^{j}(\mathbf{p})}{\partial p^{i}},
\label{theta}
\end{equation}
which is dual to the standard electromagnetic relation in position space.

In order to examine more in detail the properties of this new
field, let us consider initially the case of a constant field what
is usual in noncommutative quantum mechanics. In the case of an
harmonic oscillator expressed in terms of the original coordinates
$(\mathbf{x},\mathbf{p})$ the Hamiltonian reads
\begin{equation}
H_{\theta
}(\mathbf{x},\mathbf{p})=\frac{\mathbf{p}^{2}}{2}+\frac{k}{2}x^{2}{\text
,} \label{hamilton}
\end{equation}
from which we get: $p^{i}=m\stackrel{.}{x}^{i}-kq_{\theta }\theta ^{ij}x_{j}$
\ , $\stackrel{.}{p}^{i}=-kx^{i}$\ \ \ and the equation of motion

\begin{equation}
m\stackrel{..}{x}^{i}=kq_{\theta }\theta
^{ij}\stackrel{.}{x}_{j}-kx^{i}{\text ,}
\end{equation}
which corresponds formally to a particle in a harmonic oscillator
submitted to an external constant magnetic field. From equation
(\ref{theta}) we deduce that $a_{\theta
}^{i}(\mathbf{p})=q_{\theta }$\ $\theta ^{ij}p_{j}$, so
$X^{i}=x^{i}+\frac{1}{2}q_{\theta }\theta ^{ij}p_{j}$, and the
Hamiltonian can then be written
\begin{equation}
H_{\theta }(\mathbf{X},\mathbf{p})=\frac{\left( m_{*}^{-1}\right)
^{ij}p_{i}p_{j}}{2}+\frac{k}{2}\mathbf{X}^{2}-k\frac{q_{\theta }}{2m}%
\overrightarrow{\Theta }.\overrightarrow{\mathcal{L}},
\end{equation}
with $\theta ^{ij}=\varepsilon ^{ijk}\Theta _{k}$, $\mathcal{L}^{i}(\mathbf{X%
},\mathbf{p})=\frac{1}{2}\varepsilon ^{i}{}_{jk}\left(
X^{j}p^{k}+p^{k}X^{j}\right) $ and $\sigma ^{ij}=\delta ^{ij}\mathbf{\Theta }%
^{2}-\Theta ^{i}\Theta ^{j}$, the dual tensor of the Maxwell
constraint tensor. Note that the interaction with the field
$\theta $ is due to the presence of the position dependent
harmonic potential and leads to a dual paramagnetic interaction
which could be experimentally observable. Like in solid state
physics of an electron in the effective periodic potential of the
ions, the particle in the $\theta $ field acquires an effective
mass
tensor $m_{*}^{ij}=m\left( \delta ^{ij}+\frac{\hbar ^{2}kq_{\theta }^{2}}{4}%
\sigma ^{ij}\right) ^{-1}$ which breaks the homogeneity of space.
This strong analogy with the vacuum of the solid state leads us to
regard this field as a property of the vacuum.

Consider now the problem of angular momentum. It is obvious that the angular
momentum expressed according to the canonical coordinates satisfies the
angular momentum algebra however it is not conserved

\begin{equation}
\frac{d\overrightarrow{\mathcal{L}}(\mathbf{X},\mathbf{p})}{dt}=kq_{\theta }%
\overrightarrow{\mathcal{L}}\wedge \overrightarrow{\Theta }.
\end{equation}
In the original $\left( x,p\right) $ space the usual angular
momentum $\ L^{i}(\mathbf{x},\mathbf{p})=\varepsilon
^{i}{}_{jk}x^{j}p^{k}$ , does not satisfy this algebra. So it
seems that there are no rotation generators in the $\left(
x,p\right) $ space. We will now prove that a true angular momentum
can be defined only if $\theta $ is a non constant field.

From the definition of the angular momentum we deduce the
following commutation relations
\[
\lbrack x^{i},L^{j}]=i\hbar \varepsilon ^{ijk}x_{k}+i\hbar q_{\theta
}\varepsilon ^{j}{}_{kl}p^{l}\theta ^{ik}(\mathbf{p}),
\]
\[
\lbrack p^{i},L^{j}]=i\hbar \varepsilon ^{ijk}p_{k},
\]
\[
\lbrack L^{i},L^{j}]=i\hbar \varepsilon ^{ij}{}_{k}L^{k}+i\hbar q_{\theta
}\varepsilon ^{i}{}_{kl}\varepsilon ^{j}{}_{mn}p^{l}p^{n}\theta ^{km}(%
\mathbf{p}),
\]
showing in particular that the $sO(3)$ Algebra is broken. To restore the
angular momentum algebra consider the transformation law
\begin{equation}
L^{i}\rightarrow \Bbb{L}^{i}=L^{i}+M_{\theta }^{i}(\mathbf{x,p}),
\end{equation}
and require the usual algebra
\[
\lbrack x^{i},\Bbb{L}^{j}]=i\hbar \varepsilon ^{ijk}x_{k},
\]
\[
\lbrack p^{i},\Bbb{L}^{j}]=i\hbar \varepsilon ^{ijk}p_{k},
\]
\begin{equation}
\lbrack \Bbb{L}^{i},\Bbb{L}^{j}]=i\hbar \varepsilon ^{ijk}\Bbb{L}_{k}.
\label{lie}
\end{equation}
The second equation implies the position independent property

\begin{equation}
M_{\theta }^{j}(\mathbf{x,p})=M_{\theta }^{j}(\mathbf{p}),
\end{equation}
while the third leads to
\begin{equation}
M_{\theta }^{i}(\mathbf{p})=\frac{1}{2}q_{\theta }\varepsilon
{}_{jkl}p^{i}p^{l}\theta ^{kj}(\mathbf{p}).
\end{equation}
Putting this equation in (\ref{lie}) we are led to a dual Dirac
monopole \cite {DIRAC} defined in momentum space
\begin{equation}
\overrightarrow{\Theta }(\mathbf{p})=\frac{g_{\theta }}{4\pi }\frac{%
\overrightarrow{p}}{\mathbf{p}^{3}},  \label{monopole}
\end{equation}
where we introduced the dual magnetic charge $g_{\theta }$ associated to the
$\Theta $ field. Consequently we have

\begin{equation}
\overrightarrow{M}_{\theta }(\mathbf{p})=-\frac{q_{\theta }g_{\theta }}{4\pi
}\frac{\overrightarrow{p}}{\mathbf{p}}{\text ,}
\end{equation}
which is the dual of the famous Poincare momentum introduced in positions
space \cite{POINCARE,NOUS1}. Then the generalized angular momentum

\begin{equation}
\overrightarrow{\Bbb{L}}=m\left( \overrightarrow{r}\wedge \overrightarrow{p}%
\right) -\frac{q_{\theta }g_{\theta }}{4\pi }\frac{\overrightarrow{p}}{%
\mathbf{p}}{\text ,}
\end{equation}
is a genuine angular momentum satisfying the usual algebra. It is the
summation of the angular momentum of the particle and of the dual monopole
field. One can check that it is a conserved quantity.

The duality between the monopole in momentum space and the Dirac monopole is
due to the symmetry of the commutation relations in noncommutative quantum
mechanics where $\left[ x^{i},x^{j}\right] =i\hbar q_{\theta }\varepsilon
^{ijk}\Theta _{k}(\mathbf{p})$ and the usual quantum mechanics in a magnetic
field where $\left[ v^{i},v^{j}\right] =i\hbar q\varepsilon ^{ijk}B_{k}(%
\mathbf{x})$. Therefore the two gauge fields $\Theta (\mathbf{p})$ and $B(%
\mathbf{x})$ are dual to each other.

Note that in the presence of \ the dual monopole the Jacobi identity (\ref
{jacob}) fails:
\begin{eqnarray}
\left[x^{i},\left[ x^{j},x^{k}\right] \right] +\left[ x^{j},\left[
x^{k},x^{i}\right] \right] +\left[ x^{k},\left[ x^{i},x^{j}\right]
\right] &=& \nonumber \\
-q_{\theta }\hbar ^{2}\frac{\partial \Theta ^{i}(\mathbf{p})}{\partial p_{i}%
}=-4\pi q_{\theta }\hbar ^{2}g_{\theta }\delta ^{3}(\mathbf{p}).
\end{eqnarray}
One can interpret this by analogy with the explanation given by
Jackiw \cite{JACKIW} of a comparable violation of the Jacobi
identity between momentum by the Dirac monopole in standard
quantum mechanics: the presence of \ the monopole in momentum
space is related to the breaking of the translations group of
momentum. As a consequence the addition law of momentum is
different from the usual Galilean additional law. Indeed if we
define the
element of the translations group of momentum by $T(\mathbf{b})=\exp \left( i%
\overrightarrow{r}.\overrightarrow{b}/\hbar \right) ,$ we have the
following relation
\begin{equation}
T(\mathbf{b}_{1})T(\mathbf{b}_{2})=\exp \left\{ i\frac{q_{_{\theta }}}{\hbar
}\Phi \left( \mathbf{p};\mathbf{b}_{1},\mathbf{b}_{2}\right) \right\} T(%
\mathbf{b}_{1}+\mathbf{b}_{2}),
\end{equation}
where $\Phi \left( \mathbf{p};\mathbf{b}_{1},\mathbf{b}_{2}\right) $ is the
flux of $\Theta $ through a triangle with three tops located by the vectors
: $\overrightarrow{p}$, $\overrightarrow{p}+\overrightarrow{b_{1}}$, and $\
\overrightarrow{p}+\overrightarrow{b_{1}}+\overrightarrow{b_{2}}$ . This
term is responsible for the violation of the associativity which is only
restored if the following quantification equation is satisfied
\begin{equation}
\int d^{3}p\frac{\partial \Theta ^{i}}{\partial p_{i}}=\frac{2\pi n\hbar }{%
q_{\theta }}
\end{equation}
leading to \ $q_{\theta }g_{\theta }=\frac{n\hbar }{2},$ in
complete analogy with Dirac's quantization \cite{JACKIW}.

It is interesting to mention that singular configuration in momentum space, seems to have been discovered in the very beautiful experiments of Fang and al. \cite{FANG} 
in the context of the anomalous Hall effect in a ferromagnetic crystal. The strong analogy between this result and the monopole we deduced from symmetry consideration in noncommutative quantum mechanics, suggest us interpreting  their Berry curvature in the AHE as our noncommutative field. The main point is the consideration of the Berry phase
\[
a_{n}^{\mu }(\mathbf{k})=i\left\langle u_{n\mathbf{k}}\right| d_{k}\left|
u_{n\mathbf{k}}\right\rangle
\]
where the wave function $u_{n\mathbf{k}}\left( x\right) $ are the periodic
part of the Bloch waves. In their work, the authors introduced a gauge
covariant position operator of the wave packet associated to an electron in
the $n$ band
\begin{equation}
x^{\mu }=i\frac{\partial }{\partial k_{\mu }}-a_{n}^{\mu }(\mathbf{k}),
\end{equation}
whose commutator is given by
\begin{equation}
\left[ x^{\mu },x^{\nu }\right] =\frac{\partial a_{n}^{\nu }(\mathbf{k})}{%
\partial k^{\mu }}-\frac{\partial a_{n}^{\mu }(\mathbf{k})}{\partial k^{\nu }%
}=-iF^{\mu \nu }(\mathbf{k})  \label{comm}
\end{equation}
where $F^{\mu \nu }(\mathbf{k})$ is the Berry curvature in momentum space.

The connection with our noncommutative quantum mechanics theory is
then clearly apparent. The $\theta (\mathbf{p})$ field corresponds
to the Berry curvature $F(k)$ and $a_{\theta }(\mathbf{p})$ is
associated to the Berry phase $a_{n}(k)$. This shows that physical
situations with a Berry phase living in momentum space could be
expressed in the context of a noncommutative quantum mechanics. 
Of course this formal analogy requires more work to deepen the 
relation between the noncommutative quantum mechanics formalism and 
the Berry phase in momentum space.

Our work is justified by the will to preserve exact symmetries.
For that we found the necessity to promote the $\theta $ parameter
of the noncommutative quantum mechanics to a $\theta (\mathbf{p})$
field. Then we showed that the restoration of the Heisenberg
algebra implies the existence of a dual gauge field in momentum
space. We proved that configuration of the field which makes it
possible to build an angular momentum which satisfies the $sO(3)$
algebra and which is preserved, is a dual monopole in momentum
space. This monopole is responsible for the violation of the
Jacobi identity and implies the non associativity of the law of
addition of the momentum. To restore associativity a Dirac's
quantization of the dual charges is necessary. As a
physical realization of our theory we can interpret the $\theta (\mathbf{p}%
)$ field as a Berry curvature associated to a Berry phase expressed in
momentum space in the context of the anomalous Hall effect. 

We benefitted from conversations with Jos\'{e} Lag\`{e}s and correspondences
with Peter Horvathy.

\end{document}